\documentclass[fleqn,twoside]{article}
\usepackage{espcrc2}

\usepackage{graphicx}
\usepackage[figuresright]{rotating}
\usepackage{dcolumn}
\usepackage{bm}

\newcommand{\AmS}{{\protect\the\textfont2
  A\kern-.1667em\lower.5ex\hbox{M}\kern-.125emS}}

\newcommand{\tmpred}{red }

\newcommand{\tmpgreen}{green }

\def\chpt{\raise0.4ex\hbox{$\chi$}PT}
\def\schpt{S\raise0.4ex\hbox{$\chi$}PT}
\def\figref#1{Fig.~\ref{fig:#1}}
\def\Figref#1{Figure~\ref{fig:#1}}
\def\figrefs#1#2{Figs.~\ref{fig:#1} and \ref{fig:#2}}

\def\leftvec{\raise1.5ex\hbox{$\leftarrow$}\kern-.85em}
\def\half{{\scriptstyle \raise.2ex\hbox{${1\over2}$}}}
\def\threehalves{{\scriptstyle \raise.15ex\hbox{${3\over2}$}}}
\def\third{{\scriptstyle \raise.15ex\hbox{${1\over3}$}}}
\def\third{{\scriptstyle \raise.15ex\hbox{${1\over3}$}}}
\def\twothirds{{\scriptstyle \raise.15ex\hbox{${2\over3}$}}}
\def\fourth{{\scriptstyle \raise.15ex\hbox{${1\over4}$}}}
\def\gtwid{\raise.3ex\hbox{$>$\kern-.75em\lower1ex\hbox{$\sim$}}}
\def\ltwid{\raise.3ex\hbox{$<$\kern-.75em\lower1ex\hbox{$\sim$}}}

\def\cO{{\cal O}}

\def\prd#1{Phys.\ Rev.\ {\bf D#1}}

\def\MeV{{\rm Me\!V}}
\def\GeV{{\rm Ge\!V}}
\def\ie{{\it i.e.},\ }

\def\et{{\it et al.}}

\def\msbar{{\overline{\rm MS}}}

\title{Results for light pseudoscalars from three-flavor simulations}

\author{The MILC Collaboration: C.~Aubin,\hskip-0.03in
\address{{\vskip-0.10in{\hskip 0.07in Department of Physics, Washington
University, St.~Louis, MO 63130, USA \vskip -.1truecm}}} 
C.~Bernard,$\null^{\rm a}$ 
\hskip-0.03in\thanks{presented by C.\ Bernard at {\it Lattice 2004}, Fermilab, June 21--26, 2004}
C.~DeTar,\hskip-0.03in
\address{Physics Department, University of Utah, Salt Lake City, UT 84112, USA \vskip -.15truecm} 
Steven~Gottlieb,\hskip-0.03in
\address{Department of Physics, Indiana University, Bloomington, IN 47405, USA \vskip -.15truecm} 
E.B.\ Gregory,\hskip-0.03in
\address{Department of Physics, University of Arizona, Tucson, AZ 85721, USA \vskip -.15truecm} 
Urs~M.~Heller,\hskip-0.03in
\address{American Physical Society, One Research Road, Box 9000, Ridge, NY 11961, USA \vskip -.15truecm} 
J.E.~Hetrick,\hskip-0.03in
\address{Physics Department, University of the Pacific, Stockton, CA 95211, USA \vskip -.15truecm} 
J.\ Osborn,$\null^{\rm b}$ 
R.L.~Sugar\hskip0.005in
\address{Department of Physics, University of California, Santa Barbara, CA
93106, USA \vskip -.15truecm} 
and
D.~Toussaint$\null^{\rm d}$
\advance\baselineskip -2pt
} 
       
\begin{document}

\begin{abstract}
We compute pseudoscalar meson masses and decay
constants in partially quenched QCD with 
three dynamical flavors of improved staggered quarks. 
Fitting the lattice data to
staggered chiral perturbation theory (\schpt) forms and extrapolating in
quark mass and lattice spacing, we find:
$f_\pi  =   129.5 \pm 0.9\pm 3.5 \; \MeV$,
$ f_K  =   156.6 \pm 1.0\pm 3.6 \; \MeV $, and
$f_K/f_\pi   =  1.210(4)(13)$. 
The value for $f_K/f_\pi$ implies $|V_{us}|=0.2219(26)$.  
We also obtain $m_u/m_d = 0.43(0)(1)(8)$, or, with 
more input on electromagnetic (EM) effects from the continuum,
$m_u/m_d = 0.41(0)(1)(4)$. 
Using the one-loop perturbative mass renormalization \cite{strange-mass}, 
we determine $m_u$ and $m_d$ in the $\msbar$ scheme.
\vspace{-.2truecm}
\end{abstract}

\maketitle
The properties of the  $\pi$-$K$ system are amenable to high-precision lattice computation.
At fixed lattice spacing and quark mass, 
our improved staggered \cite{MILC_SPECTRUM2} computation 
achieves a statistical accuracy of 0.1\% to 0.4\% for decay constants
and 0.1\% to 0.7\% for squared meson masses, down to $m_\pi\!\approx\!250\, \MeV$ 
($m_\pi/m_\rho\!\approx\! 0.3$).
This allows us to check that the chiral behavior is as expected from continuum
chiral perturbation theory (\chpt) plus the controlled $\cO(a^2)$ corrections
predicted by \schpt\ \cite{SCHPT}.
The check is especially important because our implementation of staggered fermions
uses the as yet unproven $\root 4 \of {\rm Det}$ trick to move from 4 ``tastes'' (doublers)
to 1. With  \schpt, the chiral and continuum extrapolations can then be performed to
extract $f_\pi$ and $f_K$, quark masses and ratios, and several
low energy constants.  A detailed account of our methods and results appears in Ref.~\cite{FPI04}.
Here, we give an overview and update our light quark mass results to take
into account continuum analyses of EM effects.

Our computation uses seven lattice runs; five make up the ``coarse'' set ($a\approx 0.125\,{\rm fm}$),
with degenerate $u,d$ sea quark mass $a\hat m'= 0.005$ to $0.03$ and strange sea
quark mass $am'_s=0.05$, and two make up the ``fine'' set ($a\approx 0.09\,{\rm fm}$),
with $a\hat m'= 0.0062,0.0124$, and 
$am'_s=0.031$. (The primes distinguish simulation masses from their physical 
counterparts.)
For Goldstone 
meson masses and decay constants we have
extensive partially quenched data: all combinations  of 8 or 9 valence masses between 
$0.1m'_s$ ($0.14m'_s$ on the fine set) and $m'_s$.
For other tastes, we have most  of the full QCD pion masses 
but no decay constants and no partially quenched data.

Given the available data, 
we can only hope to make precise computations for Goldstone mesons.  
However, since mesons with other tastes affect the Goldstone properties through chiral loops,
we need to know the masses of those mesons to reasonable precision. 
We get the needed splittings and the
slope in quark mass from a tree-level \schpt\ fit to full
QCD data for pions of all tastes.  Using these tree-level
masses in the one-loop terms gives an NNLO error in the final results;
since the maximum error in the tree-level fit is only $\approx 7\%$, the error in the final
results turns out to be well under $1\%$, \ie negligible.

To get good fits to \schpt\ forms,  we need to
place upper limits on the valence quark masses ($m_x$, $m_y$) considered. 
 We consider 3 data subsets: subset {\it I} (94 data points) with  $m_x\!+\!m_y \le 0.40m'_s$ 
(coarse) and $\le 0.54m'_s$ (fine);  subset {\it II} (240 points) with $m_x\!+\!m_y \le 0.70m'_s$ 
(coarse) and $\le 0.80m'_s$ (fine); and subset {\it III} (416 points) with  $m_x\!+\!m_y \le 1.10m'_s$ 
(coarse) and $\le 1.14m'_s$ (fine).
 We can tolerate heavier valence masses (compared to $m'_s$) on the fine lattices, since both
$m'_s/m_s$ and the taste splittings are smaller. 

 On subset {\it I}, the maximum valence-valence Goldstone mass is $\approx\!\! 350\MeV$.
 Adding on the average taste splitting gives $\approx\!\! 500\MeV$; the worst case is
$\approx\!\! 580\MeV$.  Defining a chiral expansion parameter $\chi\!\equiv\! (500\,\MeV)^2/(8 \pi^2 f_\pi^2)$, 
we thus expect the errors of NLO \schpt\ to be of order $\chi^2\approx\! 3.5\% $. Since this is much larger
than our statistical errors, NNLO terms are clearly needed, even for this lightest
mass set.  Unfortunately, the NNLO \schpt\ logs have not been calculated.  But for the
higher masses where the NNLO terms are important,
the logs should be smoothly varying and well approximated by NNLO analytic terms. Our 
``NNLO'' fit functions therefore include the full NLO expressions 
supplemented by the analytic terms at NNLO.
We fit decay constants and masses together, including all correlations; this
constrains the new \schpt\ hairpin parameters, which are common to both.
We also fit coarse and fine lattices together, which constrains the overall 
lattice spacing dependence.

The NNLO fit has 20 unconstrained parameters:  2 at LO,  4 physical ones ($L_i$)
at NLO, 4 taste-violating ones at NLO, and 10 physical
NNLO analytic parameters.
 There are an additional 16 tightly-constrained parameters that allow for the
variation of physical parameters with $a$ ($\sim\! \alpha_Sa^2\Lambda_{\rm QCD}^2\approx 2\%$).
Finally, 4 tightly-constrained parameters allow the scale determinations to vary within
statistical errors.

\begin{figure}[t!]
\resizebox{7.5cm}{!}
{\includegraphics{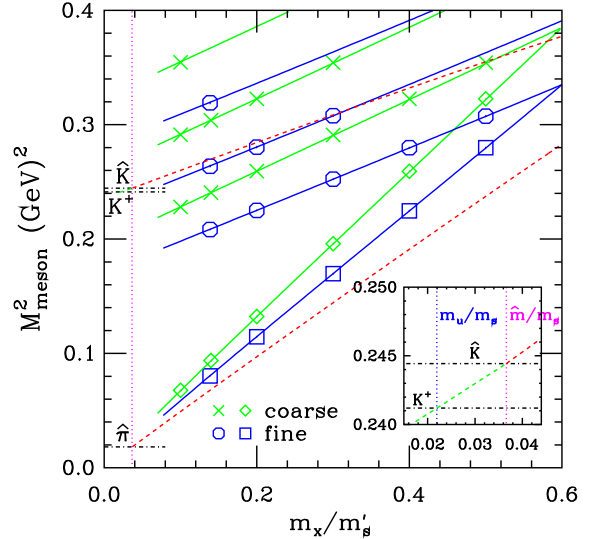}}
\vspace{-1.5cm}
\caption{
Squared meson masses {\it vs.}\ $m_x/m'_s$. 
``Kaon'' points (crosses and octagons)
have $m_y=m'_s$, $ 0.8 m'_s$, or $0.6 m'_s$; 
``pion'' points  (diamonds and squares) have $m_x=m_y$.
\label{fig:mquark}
}
\vspace{-0.6cm}
\end{figure}

\begin{figure}[t!]
\resizebox{7.5cm}{!}
{\includegraphics{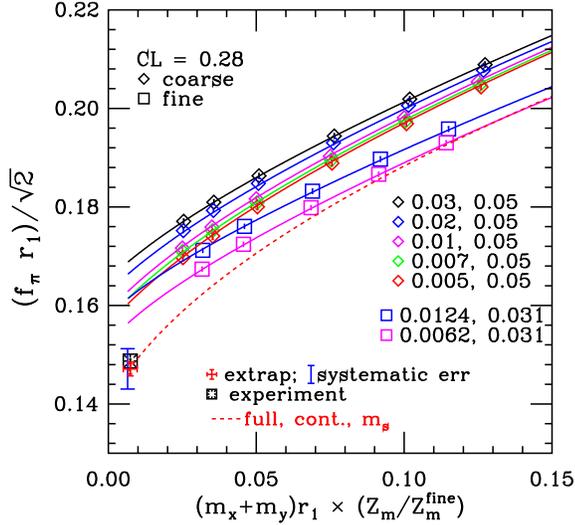}}
\vspace{-1.2cm}
\caption{
``Pion'' decay constants with $m_x=m_y$. 
\label{fig:fpi}
}
\vspace{-0.7cm}
\end{figure}

We are able to find good NNLO fits for subsets {\it I}\/ and {\it II}.
In subset {\it III}, even NNLO fits break down.  Yet it is subset {\it III}
that can allow us to interpolate around $m_s$.  The strategy for subset {\it III}
is to fix the LO and NLO terms from lower mass fits; 
then add on {\it ad hoc}\/ higher order terms to get a good interpolating 
function around $m_s$.  We use such fits for central values of
quark masses and decay constants; results 
from the other subsets are included in systematic error estimates. 
\Figref{mquark} shows the fit results in subset {\it III}\/ for partially-quenched
meson masses; while \figref{fpi} shows the same fit for pion decay constants.
Points and fit lines in these plots have been
corrected for finite volume effects at one-loop in \schpt.

To find physical values of the (bare) quark masses, we 
first extrapolate the chiral parameters to the
continuum.  The extrapolation is linear in the leading discretization errors,
either $\alpha_S^2 a^2$ (for taste-violating terms) or 
$\alpha_S a^2$ (for ``generic'' discretization effects in the physical parameters). 
We then set valence and 
sea quark masses equal (\ie go to ``full QCD''), and iteratively
interpolate in $m'_s$ and extrapolate in $\hat m'$ until
both the kaon and the pion reach
their physical masses.
The extrapolations of masses and decay constants in $\hat m'$
(after $m'_s$ has been adjusted to $m_s$) are shown as 
dashed \tmpred lines in \figrefs{mquark}{fpi}.

We must distinguish here between the experimental meson masses, QCD
masses (no electromagnetism), and those with EM effects and isospin violations
turned off, $m_{\hat K}$ and $m_{\hat \pi}$, which are the 
``physical'' masses that determine $\hat m$ and $m_s$ in our simulation.
From continuum \chpt, these masses  are related by
\begin{eqnarray*}
\label{eq:DashenViol}
		m^2_{\hat \pi} & \approx & (m_{\pi^0}^{\rm QCD})^2 \hspace{0.2truecm} \approx\hspace{0.2truecm}  m_{\pi^0}^2 \nonumber \\
		m^2_{\hat K} & \approx & \big[\,(m_{K^0}^{\rm QCD})^2
 + (m_{K^+}^{\rm QCD})^2\;\big]/2 \nonumber \\
		(m_{K^0}^{\rm QCD})^2 & \approx & m^2_{K^0} \\
		 (m_{K^+}^{\rm QCD})^2 & \approx & m^2_{K^+} -(1+\Delta_E)
\left(m_{\pi^+}^2 - m_{\pi^0}^2\right) \ ,\nonumber 
\end{eqnarray*}
where $m_{\pi^0}$, $m_{\pi^+}$, $m_{K^0}$, and
$m_{K^+}$ are the experimental masses, and $\Delta_E$ 
parametrizes the violations of Dashen's 
theorem \cite{Dashen:eg}.  A conservative choice, with minimal continuum input,
is $\Delta_E\!=\!1\pm1$.  Using more information from the
continuum gives \cite{CONTINUUM} $\Delta_E=1.2\pm0.5$. 

Once $\hat m$ and $m_s$ are known, we may continue the kaon extrapolation as a function
of $m_x$ (\tmpgreen dashed line \figref{mquark} inset) until the value
$m_{K^+}^{\rm QCD}$ is reached.  This determines $m_u$ 
up to small isospin-violating corrections, which arise because
$m_u=m_d=\hat m$ for the sea quarks.

We now extrapolate the decay constants to physical quark masses,
obtaining 
\begin{eqnarray*}\label{eq:f_results}
f_\pi & = &  129.5 \pm 0.9\pm 3.6 \; \\
f_K & = &  156.6 \pm 1.0\pm 3.8 \; \\
f_K/f_\pi  & = & 1.210(4)(13)\ , 
\end{eqnarray*}
where the errors are statistical and systematic (coming mainly
from the chiral and continuum extrapolations, and,
for dimensionful quantities, the scale error).
Using Ref.~\cite{Marciano:2004uf}, the result for $f_K/f_\pi$ implies
$|V_{us}|=0.2219(26)$.

For light quark masses at $2\,\GeV$, we obtain
\begin{eqnarray*}
m_u^\msbar&  =   1.7(0)(1)(2)(2); & 1.6(0)(1)(2)(1)\;\MeV \\
m_d^\msbar&  =  3.9(0)(1)(4)(2); & 3.9(0)(1)(4)(1) \;\MeV \\
m_u/m_d & =  0.43(0)(1)(8); & 0.41(0)(1)(4) \ , 
\end{eqnarray*}
where the first result in each line comes from
the conservative choice $\Delta_E=1\pm1$; while the
second uses $\Delta_E=1.2\pm0.5$ \cite{CONTINUUM}.
In each case the first two errors are from statistics and lattice systematics; the last,
from EM effects, and the third (for $\msbar$ masses) is an estimate of perturbative
errors \cite{strange-mass,FPI04}. 

Despite the fact that we have a large number of free parameters 
in our fits, 
we believe that the chiral behavior predicted
by \schpt\ is well tested by our analysis. The evidence for this is that good
fits are not possible with any of the following alternatives:
(a) fits to the continuum form (36 parameters),
(b) fits with all chiral logs and finite volume corrections omitted, \ie keeping 
only analytic terms (38 parameters), and (c) fits with all chiral logs omitted, but finite
volume corrections included  (38 parameters).
In subset {\it II}, all these fits have $\chi^2/{\rm d.o.f.}>3$ with ${\rm d.o.f.}\ge 202$.

\end{document}